# The Thermal Unbalance Effect Induced by a Journal Bearing in Rigid and Flexible Rotors: Experimental Analysis

Thibaud Plantegenet[a], Mihai Arghir[a], Mohamed-Amine Hassini[b], Pascal Jolly[a]

[a] Institute PPRIME, UPR CNRS 3346, Université de Poitiers, ISAE ENSMA,

86962 Chasseneuil Futuroscope, France

[b] ERMES-IMSIA, EDF Lab Paris-Saclay, 91120 Palaiseau, France

**Abstract**

The present work presents the experimental analyses of a rigid (short) and of a flexible (long) rotor subject to thermal unbalance effects. The rotors are supported by a ball bearing and by a cylindrical journal bearing. The differential heating generated in the journal bearing is responsible for the thermal unbalance. The results obtained with the short rotor at 7 krpm showed an increase of the synchronous amplitudes but slight phase changes before stabilization. The pronounced hysteresis of the synchronous amplitudes obtained during coast-down proved that the amplitude increase is due to the thermal unbalance. The results obtained for the long rotor at 6.6 krpm showed the same stabilized response when start-up was performed in 180 s. However, an instability leading to journal bearing contact was triggered when the start-up time was decreased to 80 s. The presented experimental results are the synchronous amplitudes and phases, the mean temperatures and the maximum temperature differences of the journal and of the bearing and the phase lags between the high spot and the hot spot.

## Introduction

The thermal unbalance effect in rotor dynamics is not yet well predicted because the existing numerical tools have not been validated by data detected in the field due to limited access to industrial testing. The effect occurs due to the coupling of the heat generated in dynamic seals or lubricated bearings with rotordynamics. One of its characteristics is that it needs a long time to develop toward dangerous, high amplitude synchronous vibrations. Indeed, synchronous vibrations have the time scale of the rotation speed while their amplitude and phase are driven by the thermal unbalance governed by the much lower time scale of the rotor heat transfer.

The source of heat generation is different in dynamic annular seals or lubricated journal bearings. In seals, the synchronously whirling rotor may slightly rub against the labyrinth teeth or against the bristles of the brush seals. Heat is then generated in the contact point. Because the rotor whirls synchronously, heat generation is located at the same "hot spot" of the rotor. The non-uniform temperature variation around the circumference of the rotor produces a thermal bow that may increase the original mechanical unbalance. Therefore, the amplitudes of the synchronous vibration will slowly increase with continuously varying phase and the vibration vector will describe a more or less pronounced spiral. This is usually called the Newkirk effect and detailed analyses can be found in rotordynamic textbooks [1]-[4].

In 1987, Schmied [5] mentioned that a very similar phenomenon might appear in lubricated journal bearings. This time there is no light contact between the rotor and the bearing. The rotor point where the lubricant film thickness is minimum is called the "high spot". Heat is generated by the lubricant shear stresses and the heat flux is not uniform around the rotor circumference. The hot spot is located before the high spot and produces the same effect as a light contact. As for the Newkirk effect, the rotor thermal bow modifies the unbalance and the synchronous vibration vector will describe a spiral. However, because it occurs due to heat generated in lubricated bearings, it is called the Morton effect.

This instability was mentioned in the pioneering work of Morton [6] in 1975 and of Hesseborn [7] in 1978. However, the effect received little attention because it is difficult to diagnose and these references were internal reports. Clear experimental evidences were scarce and therefore theoretical predictions were not developed. The Morton effect came under scrutiny only during the 1990s.

In 1996, de Jongh and Morton [8] studied a centrifugal compressor guided by two tilting pad bearings and subject to a hysteretic behavior and spiral vibrations. The compressor was built to operate at 12 krpm far from the critical speed located at 14.5 krpm ($2^{nd}$ bending mode). However, during run-up synchronous vibrations began to increase between 10 krpm and 11.5 krpm. At 11.5 krpm the amplitudes continued to grow rapidly up to 50% of the bearing clearance while the phase was continuously changing. The shaft speed was reduced to 10 krpm and after few minutes, vibrations returned to their initial values. After rebalancing, refitting and without oil seals, the compressor showed the same instability. Further runs with light overhung weights showed that the speed at which the instability occurred became 10% higher than the operating speed. In order to understand the source of the instability, de Jongh and Morton developed a test rotor with the same dynamic characteristics as the compressor and with four temperature sensors mounted in the journal. The test rotor showed the same behavior and the temperature sensors proved that the journal bearing was the source of synchronous instability. At 10.5 krpm, the rotor temperature difference was 3°C, the phase lag between the hot spot and the high spot was 20° and the vibration amplitudes increased but remained stable. At 11.5 krpm, the differential temperature was 10°C and the spiral vibrations became unstable.

In 1998, de Jongh and van der Hoeven [9] published other results of synchronous instabilities. A centrifugal compressor showed hysteresis loops during run-up and coast-down and cyclic vibrations at constant speed, i.e. periodically increasing and decreasing synchronous amplitudes accompanied by a continuously phase change. The compressor was dismounted and a disk replaced the impeller to test the influence of the overhung weight, with and without seals. Vibrations remained low and stable for all configurations. However, a faster run-up conducted a few days later showed instabilities. Other tests conducted the same day showed stable, low amplitude vibrations. Next morning, the same test was conducted and showed again instabilities while all the other tests of the day were stable. The authors concluded that the instability occurred due to the cold morning temperature of the system. Due to the quick start-up, the journal bearing heats, but the clearance is constrained by the cold temperature of the housing. To cure the problem, a heat barrier introducing an air layer between the shaft and the sleeve was developed and successfully tested.

In 1999, Berot and Dourlens [10] presented the results of a centrifugal compressor with a large overhung impeller guided by two tilting pad bearings and a thrust bearing. The authors tested five compressors with the same architecture but with different overhung impellers. All compressors showed instabilities close to the operating speed. Some compressors showed variations of the synchronous amplitudes following slow cycles of the order of minutes with a continuous phase shift. Other tests showed rapidly increasing synchronous vibrations necessitating the shut-down of the compressor. The whirl orbits were almost centered and circular. The authors concluded that the differential heating of the journal caused the spiral vibrations. Nothing indicated a rubbing between the journal and the stator, so the differential heating could come only from the journal bearing. The authors tested other configurations with different impellers, oil supply temperatures, bearing preloads and clearances and lighter couplings, but nothing suppressed the unstable behavior. After numerical investigations, they found that reducing the L/D ratio of the bearing increases the eccentricity and changes the orbit to an elliptical one. Tests showed that the new bearings cured the instability problem.

In 2007, Marscher and Illis [11] presented a problem of an integrally geared air separation compressor supported by tilting pad bearings. The compressor operated normally for years but an improvement of the oil supply system lead to a decrease of the feeding temperature and resulted in high cyclic vibrations with constant phase shifting. The issue was resolved by increasing the inlet temperature.

Schmied et al. [12] published in 2008 an experimental analysis of a turboexpander with overhung disks. Close to the critical speed, the rotor showed high amplitude, spiral synchronous vibrations. The amplitudes remained high and followed a hysteresis when the speed was decreased thus showing that the culprit was a thermal effect. The solution for currying the instability was based on the reduction of the bearing width (as in [10]) and of the oil viscosity.

In 2011, Lorenz and Murphy [13] investigated a rotor with a large overhung disk. The amplitudes of synchronous vibrations increased very slowly during two hours. After this time interval they became excessive very rapidly thus the machine was coasted down. The authors thoroughly described the Morton effect and introduced three different responses:
- stable vibrations: the amplitude increases and the phase vary but they tend toward constant values,
- the limit cycle: the phase changes constantly but the amplitude remains constant or show cyclic increase and decrease,
- unstable vibrations: the phase changes constantly and the amplitude increases; the synchronous vibration vector describes a spiral.

In 2015, Panara et al. [14] presented a test rig for verifying theoretical predictions. The rotor had an overhung disk. Experimental results showed that the onset speed of Morton effect decreased with increasing the mass of the overhung disk. However, the rotor could become stable again once the rotation speed was slightly beyond the critical speed.

Tong and Palazzollo [15] presented in 2018 a particular test rig aimed to measure only the journal temperature difference of a synchronously whirling rotor. Two ball bearings supported the rotor at each side and a tilting pad journal bearing was located in the mid-plane. The rotor was machined with an eccentricity relative to the ball bearings centerline. The eccentricity was 32% of the radial clearance and reproduced a circular vibration orbit in the tilting pad bearing; 20 RTDs equally spaced sensors were mounted on the journal. The main conclusions of the tests were that the journal temperature difference increases almost linearly with the rotor speed and the phase lag between hot spot and high spot was speed dependent, comprised between 20°@1.2 krpm and 40°@5.5 krpm.

The results presented in [12], [13] and [14] show all the symptoms of instabilities induced by thermal unbalance: the instability is first hidden at constant speed and appears only after a certain time, it is accompanied by a clear hysteresis of the vibration amplitudes during coast-down and it may disappear if operating beyond the critical speed. Other experimental evidences of the Morton effect are discussed in [2]. Detailed presentations can also be found in three recent review articles [16], [17] and [18]. This abundant literature shows that the Morton effect is receiving an increasing attention at this moment.

The goal of the present work is to present the experimental analyses of a rigid (short) and of a flexible (long) rotor subject to thermal unbalance effects. The results obtained with the short rotor at 7 krpm showed a strong increase of the synchronous amplitudes but a slight phase change before stabilization. The pronounced hysteresis of the synchronous amplitudes obtained during coast-down shows that thermal unbalance is responsible for this result. This corresponds to a "stable" Morton effect. The results obtained for the long rotor at 6.6 krpm showed first the same "stable" Morton effect. However, a thermal unbalance instability was triggered when the start-up time was decreased from 180 s to 80 s. These experimental results are detailed and commented in the following.

## Test rig description

The test rig is illustrated in Figure 1. It consists of a hollow rotor guided by a ball bearing at its left end and by a journal bearing close to its right end. The first rotor is 430 mm long, its outer and inner diameters being 45 mm and 35 mm. The distance between the bearing centerlines is 198.5 mm. A second rotor of 700 mm long will be presented in a further paragraph.

A 1.8 kW DC Motor and a flexible coupling drive the rotor. The maximum attainable speed is 10 krpm. A slip ring is mounted at the rotor right end. The rotor mass (including the hollow shaft, one half of the coupling and the slip ring) is 2.2 kg. Three disks of 0.7 kg are overhung mounted at the right end. The disks have 12 threaded holes, equally spaced in the circumferential direction for adding unbalance masses. The static loads supported by the journal and the ball bearings were 24.5 N and 4.5 N, respectively. A Campbell diagram calculated by using the short bearing rotordynamic coefficients and a stiffness of $3 \cdot 10^8$ N/m for the ball bearing showed a rigid mode close to 60 rpm while the first elastic mode was far beyond the maximum rotation speed.

The 430 mm rotor was balanced in class G1@10krpm but the test rig required repeated dismounting of the disks and of the slip ring. Therefore, a quite large residual unbalance was present. Comparisons with theoretical predictions enabled to estimate a residual unbalance of 45 g.mm @150° on the disks.

### Journal bearing design

The journal bearing is of simple cylindrical type with an axial feeding groove located at its top (Figure 2). The bearing was made of Teflon incrusted bronze and was mounted in a steel housing. Its radial clearance is 52.2 µm and its length is 15 mm. The journal bearing is lubricated with an ISO VG 32 turbine oil (the kinematic viscosity is 25.1 cP@40°C and 11.5 cP@60°C) at ambient (room) temperature. The feeding temperature was not controlled. However, the large volume of lubricant in the tank secured a quasi-constant feeding temperature. For example, in a 4 hours test at 7 krpm rotation speed, the temperature of the tank showed a 2°-3°C increase.

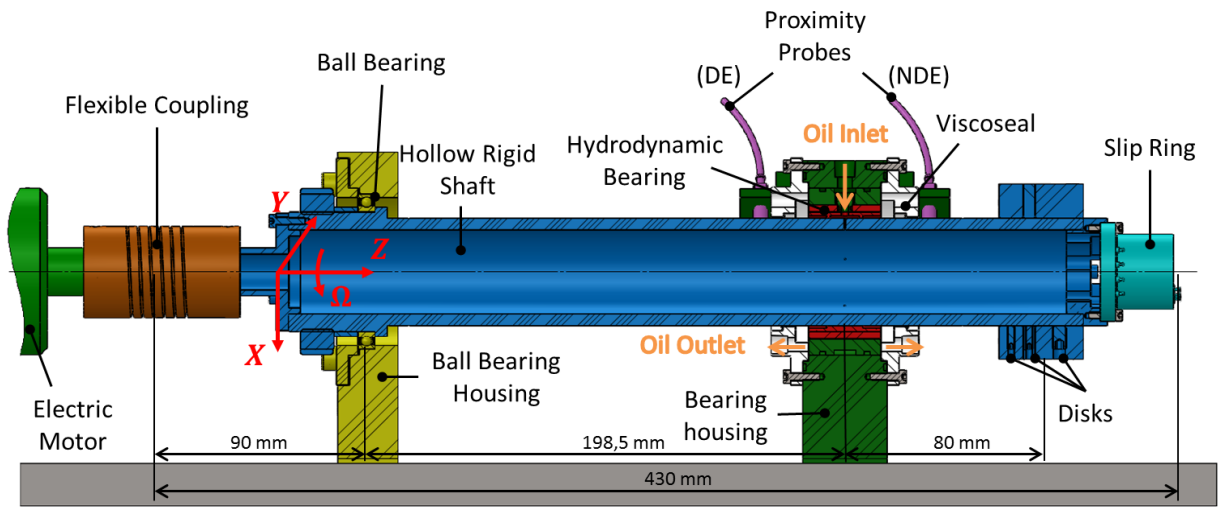

Figure 1 : The test rig equipped with the short (rigid) rotor

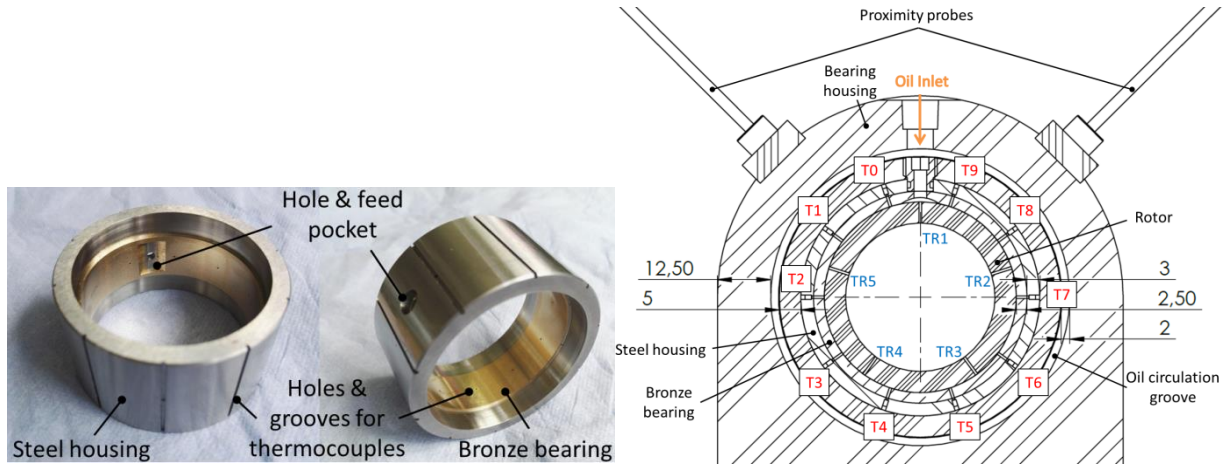

Figure 2 : The cylindrical journal bearing and its instrumentation

**Instrumentation**

The test rig is instrumented with ten thermocouples mounted in the mid-plane of the bearing and five thermocouples mounted in the journal and connected to a slip ring. The journal thermocouples are also situated in the bearing mid-plane. The thermocouples are equally spaced in the circumferential direction. All thermocouples are of T type with 0.5°C accuracy. The displacements are measured by inductive proximity probes orthogonally mounted two by two in the drive end (DE) and the non drive end (NDE) plane (Figure 1, 2). An optical sensor is used to measure the rotation speed and the phase reference. The supply pressure and temperature are also measured at the bearing inlet.

## Experimental results for the short (rigid) rotor

The first experiments showed that the rotor-bearing system was prone to sub-synchronous instabilities when approaching a rotation speed of 3 krpm. This was not considered a problem because only large amplitude synchronous vibrations were sought for the present tests. Therefore, an additional unbalance of 60.6 g.mm@180° relative to the zero-phase reference was mounted on the overhung disks. The cumulated unbalance of the complete rotor approached 102 g.mm @167°.

This unbalance stabilized the rotor vibrations when approaching 3 krpm and beyond. Indeed, the vibrations close to 3 krpm enlighten the $\Omega/2$ whirl phenomena (i.e. a linearly unstable operating regime) that changes rapidly to whip if the amplitudes are small and the speed is increased beyond the stability threshold (the self-sustained vibrations become unstable). This is not the case when the rotor has large amplitude non-linear vibrations due to important unbalance. The whirl phenomena still appears but the journal bearing may develop enough damping for overriding the whirl regime. The rotor may become again linearly stable at higher speeds. This behavior is not new in rotordynamics and it is exactly what happened in the present case.

The vibrations of the rotor-bearing system were then investigated by slowly increasing the speed from 0.6 krpm to 7.2 krpm. The bearing feeding pressure was set to 0.1 bar when the rotor was not spinning and the feeding temperature was at its ambient value (20°C).

Figure 3 shows the cascade plot of the X direction vibrations measured in the NDE plane. The 1X and the 2X components are clearly distinguishable together with a 0.5X component. The later appears at 3 krpm, attains its maximum amplitude at 5 krpm (the amplitudes of the 0.5X and 1X components are equal at this speed) and completely vanishes at 6.5 krpm. The amplitude of the 2X component remains small. The goal of the experiments was to obtain high amplitude synchronous vibrations. Therefore, systematic tests were performed only at 4 krpm and 7 krpm, outside the zone with noticeable subsynchronous vibrations.

The first experimental campaign consisted of long tests at 4 krpm. The added unbalance, the feeding pressure and temperature were not modified. These results showed no noticeable thermally induced unbalance effect and were therefore relegated to the Appendix 1. Systematic measurements were then performed for 7 krpm rotation speed.

## Synchronous amplitudes and phases at 7 krpm

Many tests with durations comprised between 30 min and 240 min were conducted at 7 krpm. The added unbalance and the feeding pressure were unmodified. The feeding temperature was comprised between 18°C and 25°C, slightly different at each test.

The results of all tests are similar, so only a 2 hours test is thoroughly presented in the following. The accuracy and repeatability of measurements are shown in Appendix 2. Displacements at the DE and the NDE planes were recorded for 1 second every 30 seconds with an acquisition frequency of 2048 Hz.

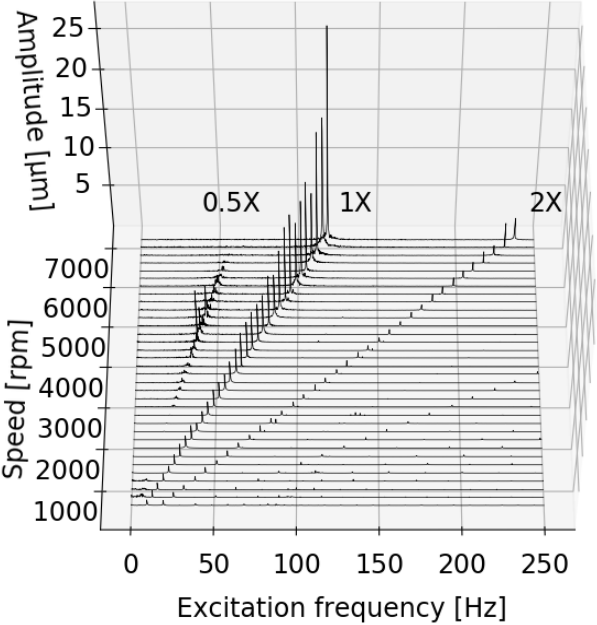

Figure 3 : Cascade plot of the short rotor measured by the NDE front sensor

Figure 4 depicts the synchronous amplitudes and phases over a long time interval. The time origin is the moment when the rotation speed of 7 krpm was attained. The synchronous amplitudes increase from 14 µm in the DE plane, and 19 µm in the NDE plane to values comprised between 26 µm and 30 µm for the DE plane and between 34 µm and 40 µm for the NDE plane. The amplitudes measured along the Y axis (horizontal) are always larger than X axis (vertical) amplitudes. In the same time, the phases of X synchronous amplitudes show an increase of 20° in both the DE and the NDE planes while the Y synchronous phases show an increase of 15°. Rapid variations occur only at the beginning of the test. This is rigorously true for the phases while the amplitudes continue to show a slight increase even after a long time interval.

Figure 5 depicts a polar plot of the synchronous amplitudes and phases. The vibrations vectors rotate counter-clockwise, in the direction of the rotation speed, with an angular extent of 20° and 15° degrees.

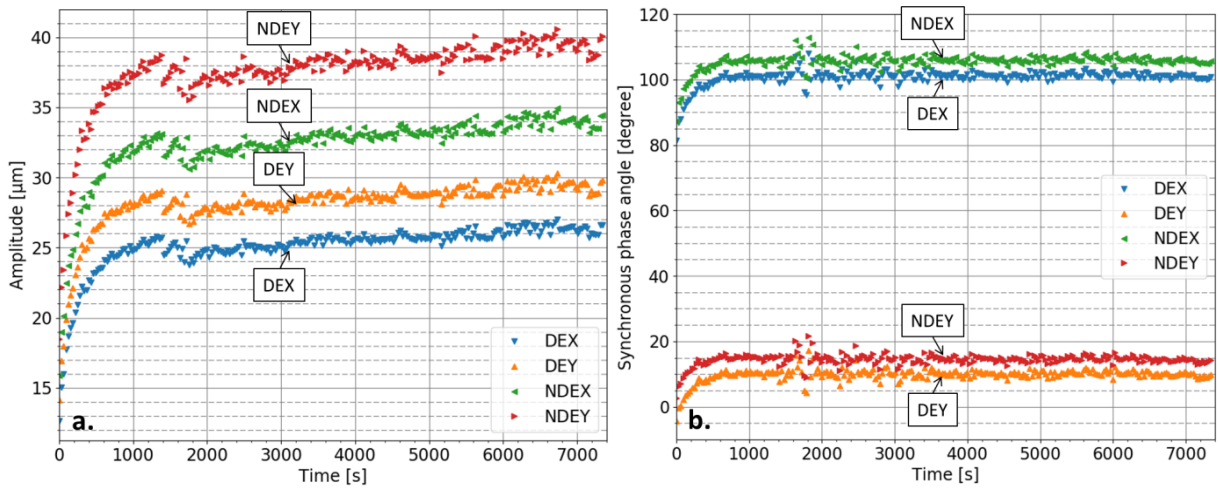

Figure 4 : Amplitudes (a) and phases (b) of synchronous vibrations of the short rotor in DE and NDE planes at 7 krpm

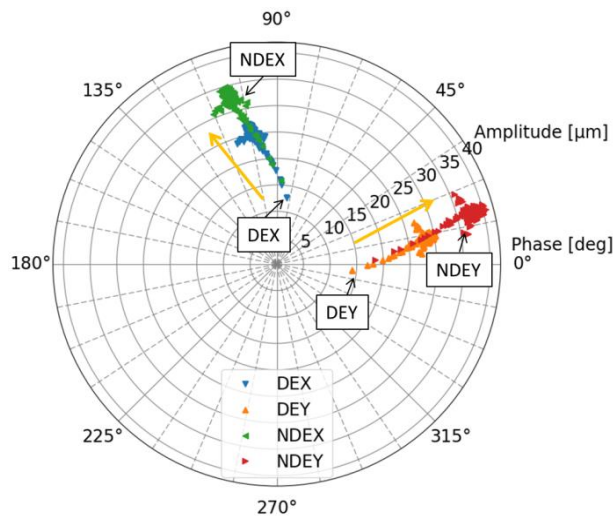

Figure 5 : Polar plot of the synchronous vibration vectors of the short rotor in the DE and the NDE planes at 7 krpm

Figure 6 depicts the circumferentially averaged values and the peak-to-peak temperature differences measured in the mid-plane of the bearing and on the journal. The rotor temperature difference is quite large, increasing from 5.5°C to 12.5°C, while the bearing temperature difference decreases from 4°C to 2.5°. Similar to the synchronous displacements, the temperature differences show major variations during the first 1400 sec of the test and then become asymptotically stable.

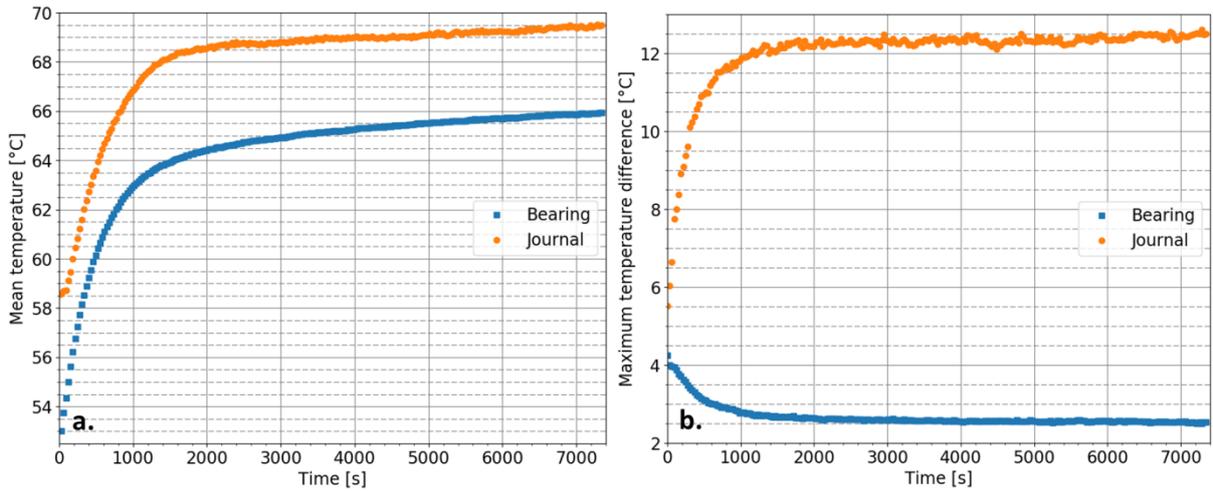

Figure 6 : Circumferentially averaged values (a) and peak-to-peak temperature differences (b) measured in the mid-plane of the bearing and on the journal at 7 krpm

The results depicted in Figure 7 show the rotor temperature difference versus the amplitude of synchronous vibrations. The R² coefficient of the linear correlation is 0.91 during the whole test and 0.99 for the first 1400 seconds. This proves that the two effects are well correlated, i.e. the increase of the synchronous amplitudes is due to the temperature difference of the journal. However, it is possible that different operating conditions will lead to other kind of correlations [7].

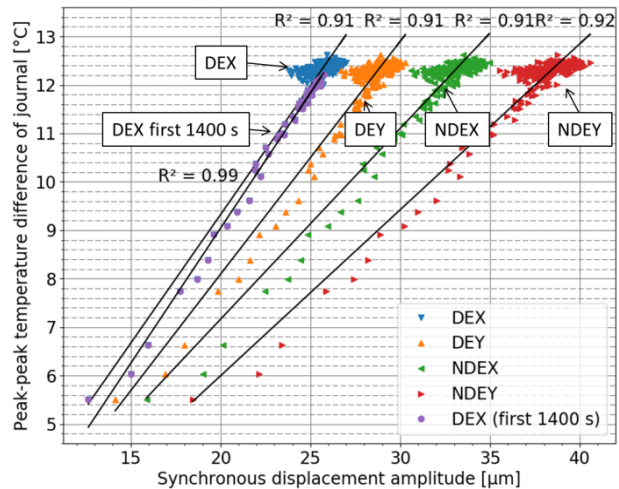

Figure 7 : Correlation between journal maximum temperature difference and synchronous displacement amplitude

### Phase lag between the hot spot and high spot at 7 krpm

The circumferential positions of the journal hot spot and high spot indicate the highest temperature and the minimum film thickness, respectively. The time variation of the phase lag between the hot spot and the high spot requires special attention for identifying the typical signature of a thermally induced unbalance effect.

The recorded rotor temperature variation is depicted in Figure 8a. The angular position is relative to the rotor zero-phase reference, i.e. in a rotating reference frame. A cubic polynomial was used to interpolate the temperatures recorded by the five rotor mounted probes during at least three rotation periods. The angular position of the rotor hot spot relative to the zero-phase reference $\Phi_{Ho/KEY}$ is depicted in Figure 8b.

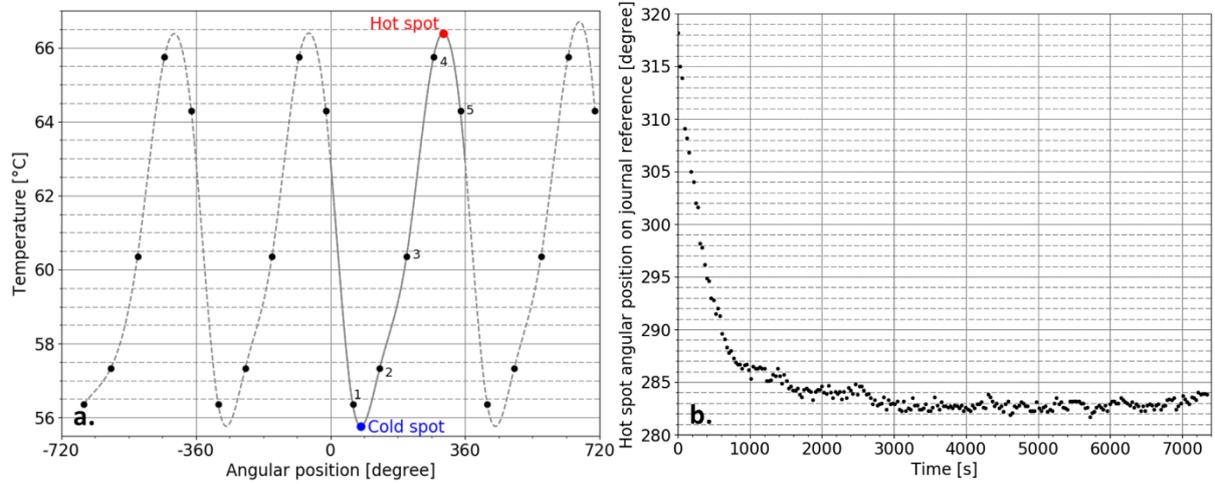

Figure 8 : Journal temperature at time = 430 s (a) and angular position of the rotor hot spot relative to the zero-phase reference (b) at Ω=7 krpm

The position of the high spot is more difficult to obtain because in the general case the rotor orbit is not centered, nor circular. Therefore, it requires at each time step the following procedure (Figure 9):

- The bearing center $O_b$ and the effective clearance in the DE and NDE planes are determined at zero rotation speed by using the proximity probes. The rotor is manually displaced up to bearing contacts.

- The synchronous orbit is constructed by using the bearing center previously determined and the information of the proximity probes mounted on the bearing. The orbit is in the fixed reference frame. The orbit in the mid plane of the journal bearing is linearly interpolated from measurements in the DE and NDE planes. The journal describes this orbit with a synchronous speed , $\omega = \Omega$ and its angular position is $\theta = \omega t$.

- The zero-phase reference point belongs to this orbit and its angular position relative to the X axis is measured by the X proximity probe, $\Phi_{KEY/X}(\theta)$.

- The location of the high spot relative to the zero-phase reference point is obtained as shown in Figure 9; $\overrightarrow{O_bO_r}$ and $\overrightarrow{O_oO_r}$ are position vectors of the rotor relative to the bearing and orbit center, respectively. They both rotate and $\overrightarrow{O_oO_r}$ has the angular speed $\omega$. The angle $\theta_b(\theta)$ is measured between $\overrightarrow{O_bO_r}(\theta)$ and the X axis. This angle measures the position of the rotor in the bearing reference frame and relative to the X axis. It also corresponds to the high spot because, as shown in Figure 9, $\overrightarrow{O_bO_r}(\theta)$ points in the direction of the minimum distance between the synchronous orbit and the bearing circumference.

- The phase difference between the high spot and the zero-phase reference point, $\Phi_{Hi/KEY}(\theta)$ is then calculated from the two previous information, $\Phi_{Hi/KEY}(\theta) = \theta_b(\theta) - \Phi_{KEY/X}(\theta)$.

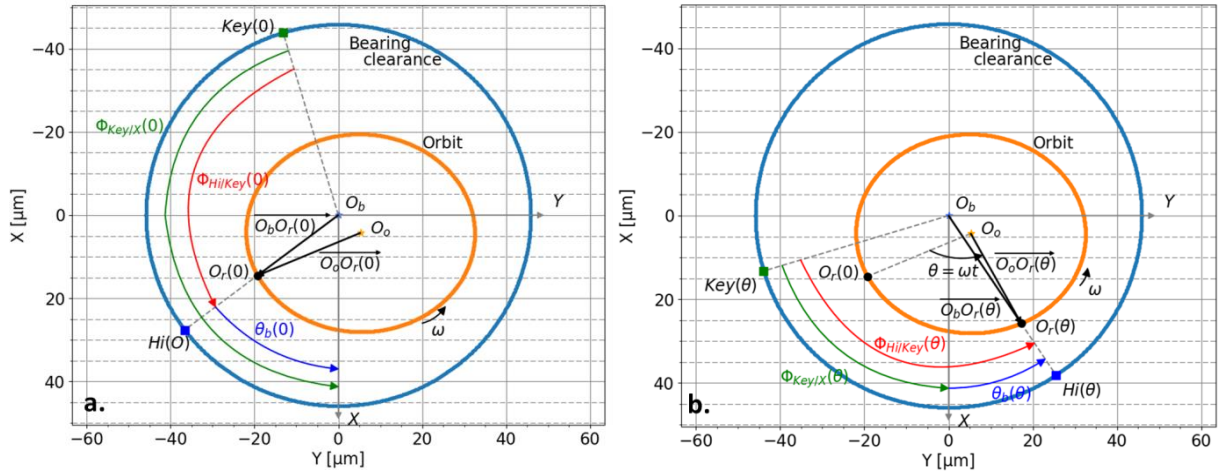

Figure 9 : Typical rotor orbit constructed from experimental data for $\theta = 0$ (a), for arbitrary $\theta = \omega t$ (b)

- An example of the estimated $\Phi_{Hi/KEY}(\theta)$ is depicted in Figure 10. The phase difference between the high spot and the zero-phase reference point is not constant over a rotation period because the orbit is not centered. Therefore, the angular speed of the vector $\overrightarrow{O_b O_r}$ identifying the position of the rotor in the bearing reference system is not equal to the precession speed $\omega$ of the vector $\overrightarrow{O_o O_r}$. The two would be rigorously equal if the orbit is perfectly centered. A mean value and a standard deviation of the phase difference between the high spot and the zero-phase reference point can then be defined for a general, off-centered orbit. These values are depicted in Figure 10 for the orbits reconstructed in the mid-plane.

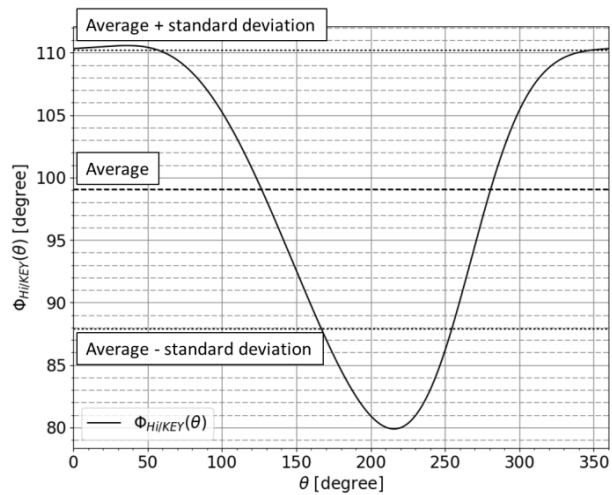

Figure 10 : Variation of the high spot phase along the orbit (results taken at t=430 s for the short rotor at 7 krpm)

- The phase lag between hot spot and high spot is finally obtained by subtracting the mean angular position of the high spot from the angular position of the hot spot, both relative to the zero-phase reference.

$$\Phi_{Ho/Hi} = \Phi_{Ho/KEY} - \overline{\Phi}_{Hi/KEY}$$

This calculation procedure is repeated at every moment when data are recorded.

The phase lag between the hot spot and the high spot is depicted in Figure 11. The error bars depicted for the values calculated in journal bearing mid-plane (Figure 11b) show the standard deviation resulting from the position of the high spot. The phase lag in the mid-plane decreases from 46° to 26°. As for the rotor temperature differences and for the synchronous amplitudes, the major part of the variations occurs in the first 1400 sec. These results confirm the values reported in the literature and show that the phase lag may vary in time.

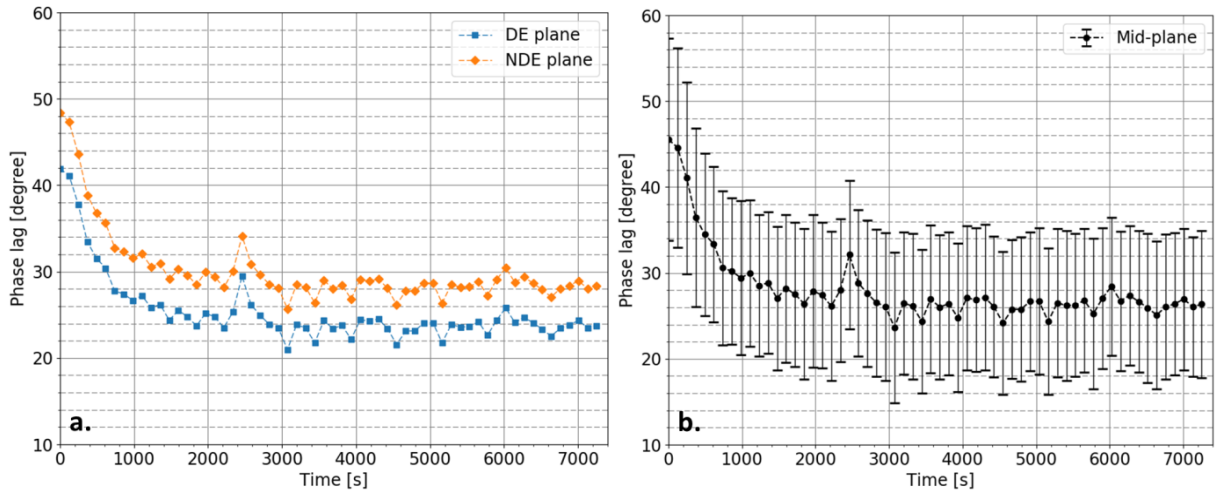

Figure 11 : Phase lag between hot spot and high spot in the DE and NDE planes (a) and in the mid plane (b) for the short rotor at 7 krpm

## Run-up/Coast-down tests

Run-up and coast down tests were subsequently performed for additionally verifying that the measured variations were due to a thermally induced unbalance. These tests were conducted with an unbalance of 143g.mm@171°. The rotation speed was progressively increased from 0 to 7 krpm in 50 seconds, kept constant for 180 seconds and then progressively decreased. The synchronous displacements and phases versus rotor speed are depicted in Figure 12. Both the synchronous amplitudes and their phases show a clear hysteresis in the zone comprised between 5.8 krpm and 7 krpm.[1] This is a typical signature of a thermally induced dynamic effect. It shows that the time scale of the thermal unbalance is much larger than the time scale of the dynamic effects characterized in the present case by the rotation speed. Journal and bearing maximum temperature differences depicted in Figure 13 show the same hysteresis effect.

---

[1] Sub synchronous vibrations were present between 3.2 krpm and 5.8 krpm.

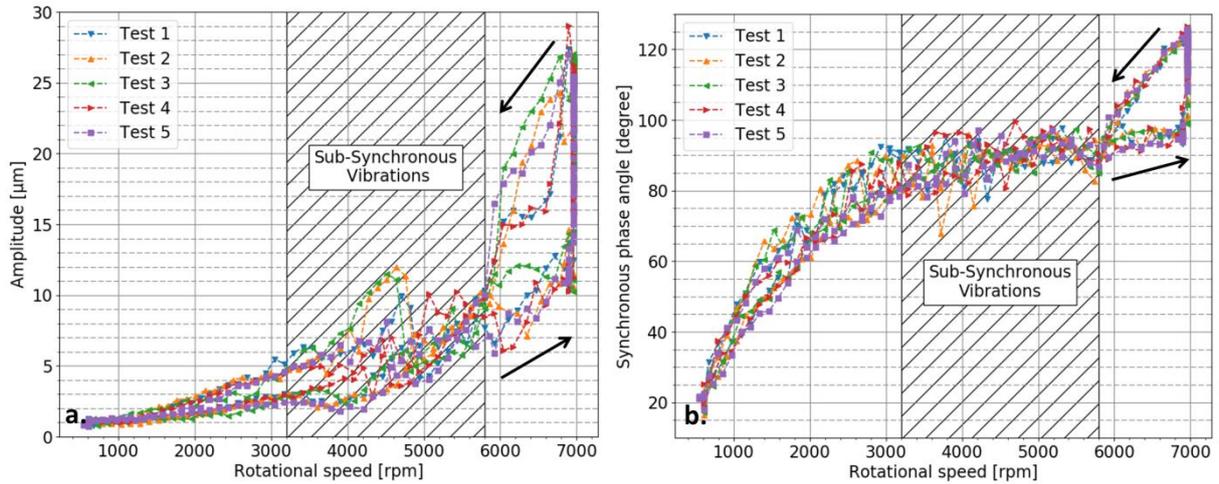

Figure 12 : Synchronous displacements (a) and phases (b) versus rotor speed for the short rotor

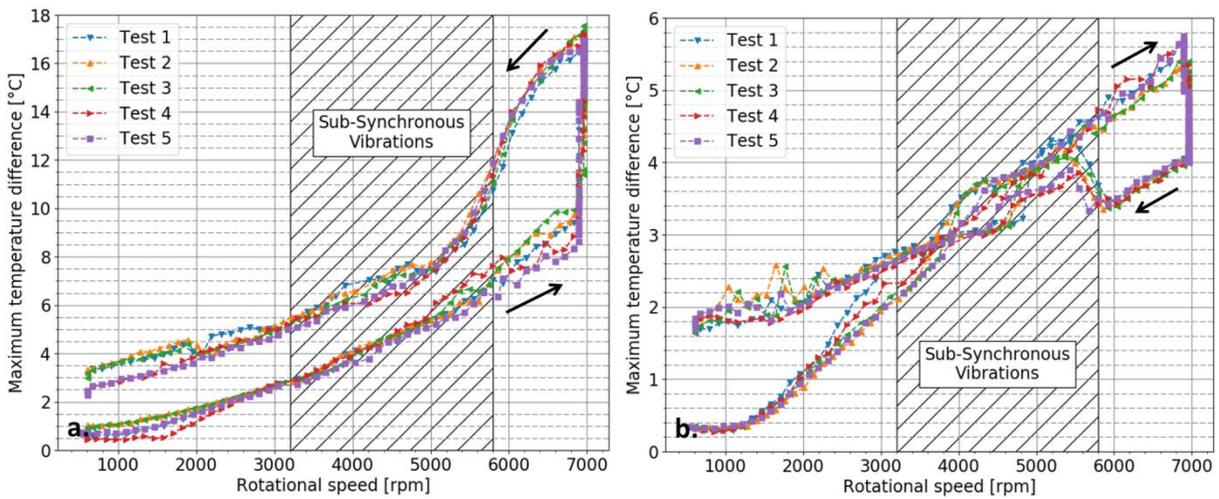

Figure 13 : Maximum journal (a) and bearing (b) temperature differences versus rotor speed for the short rotor

## Experimental results for the long (flexible) rotor

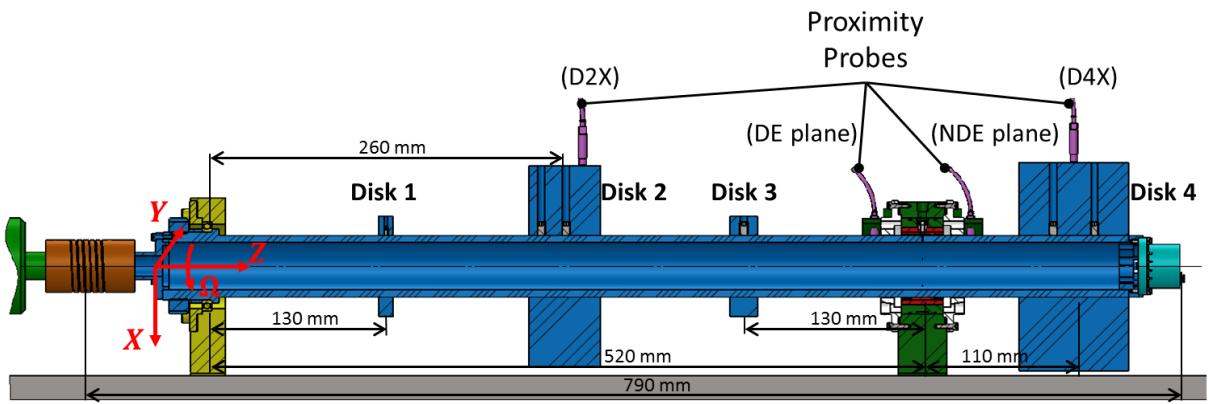

Figure 14 : The test rig equipped with the long (flexible) rotor

A second rotor of 700 mm long was designed. The rotor has the same inner and outer diameter and uses the same spindle, elastic coupling, ball and journal bearings as the short rotor. The distance between bearings is 520 mm. The instrumentation was also the same namely, two by two orthogonally mounted proximity probes at the two ends of the journal bearing, a slip ring with five thermocouples mounted at the mid-plane of the journal and ten thermocouples mounted in the mid-plane of the bearing surface. The rotor was equipped with two heavy disks, one of 6.4 kg mounted between the bearings and another of 10.4 kg overhung mounted (disk 2 and 4 in *Figure 14*, respectively). Two small disks, previously used for the short rotor were installed between the central disk and the bearings (disk 1 and 3 in *Figure 14*). Their inertia characteristics are negligible compared to the heavy disks but they serve as additional balancing planes. The characteristics of the disks are given in Table 1. Two inductive proximity transducers measure the X displacements of the two heavy disks. The results for the long rotor were recorded every second.

The Campbel diagram showed a flexible backward mode at 6.6 krpm and flexible forward mode at 8.5 krpm.

Balancing the rotor close to the envisaged operating speed of 7 krpm encountered difficulties. The rotor was therefore balanced at 4 krpm by using the four balancing planes represented by the disks and following the procedure given in [19]. After balancing, the synchronous amplitudes at 4 krpm were 5 µm and 3 µm in the DE and NDE measurement planes, respectively.

**Stable thermal response of the long rotor (test 1)**

The speed was increased to 6.6 krpm in 180 s and was then held constant. Due to the important static load on the journal bearing (≈175 N), subsynchronous vibrations were not recorded between 0 rpm and this speed. It was verified on full spectrum diagrams that all measured orbits were forward precessions. The synchronous amplitudes and phases recorded at 6.6 krpm are depicted in Figure 15. The amplitudes show a rapid increase at the beginning of the time interval but then tend asymptotically to constant values. Phases also become constant. This is the signature of the stable Morton effect also enlightened by the short rotor.

The X direction amplitudes and phases recorded in the NDE plane versus rotor speed are depicted in Figure 16. They show a clear hysteresis effect. This proves well that the amplitude increase recorded at 6.6 krpm is due to heat transfer.

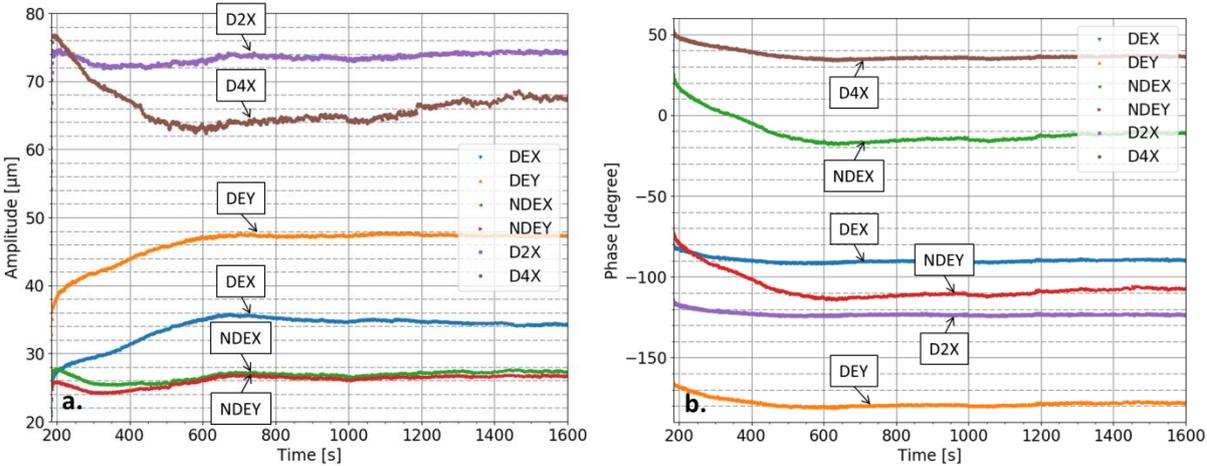

Figure 15 : Synchronous amplitudes (a) and phases (b) measured for the long rotor at 6.6 krpm, stable response of test 1

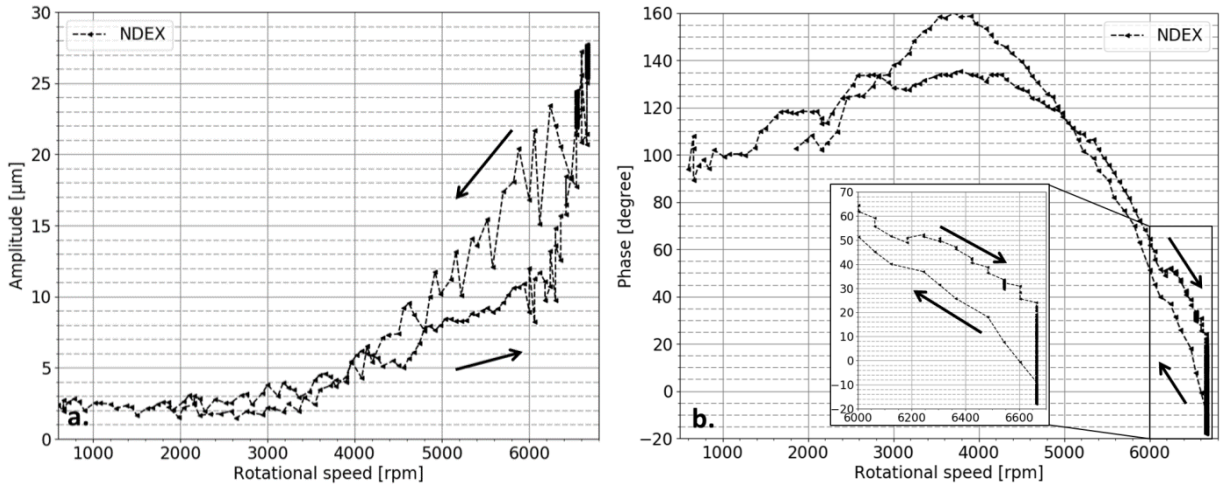

Figure 16 : Synchronous amplitudes (a) and phases (b) at NDEX recorded during start-up and coast-down of the flexible rotor, stable response of test 1

The polar plot of the vibration vectors are depicted in Figure 17. All vibration vectors turn clockwise, i.e. opposite to the rotation speed. However, the spirals are quite different from one measurement plane to another. The rotation of the vibration vector increases progressively from the plane of disk 2 to the plane of disk 4. Thus, the rotations in the DE plane are lower than in the NDE plane and both are comprised between the rotations measured in the planes of disk 2 and disk 4, respectively.

The average and the maximum temperature differences extracted from these measurements are depicted in Figure 18. All temperatures become stabilized after 600 s. The stabilized average temperatures of the journal and of the bearing are different, 62°C and 55.5°C, respectively, while the temperature differences are 14°C and 14.5°C. The peculiar variation of the maximum temperature difference of the bearing between 180 s and 600 s is due the thermal inertia of the bearing at the end of the start-up process. The sudden jump of the temperature difference around 550s is due to measurement uncertainty of the lowest bearing temperature probe.

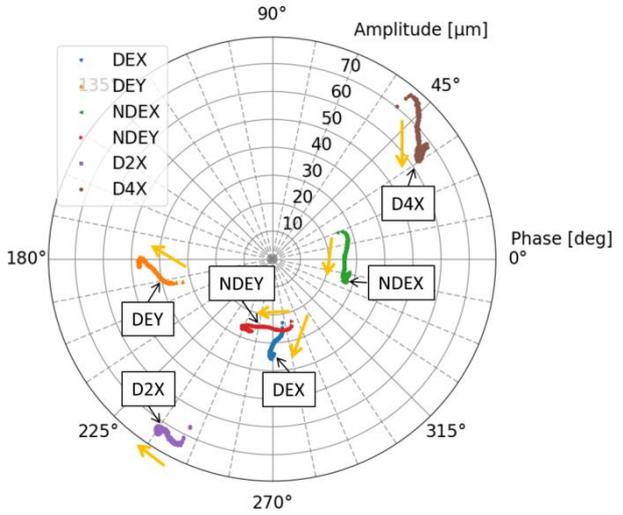

Figure 17 : Polar plot of the vibration vector measured for the long rotor at 6.6 krpm, stable response of test 1

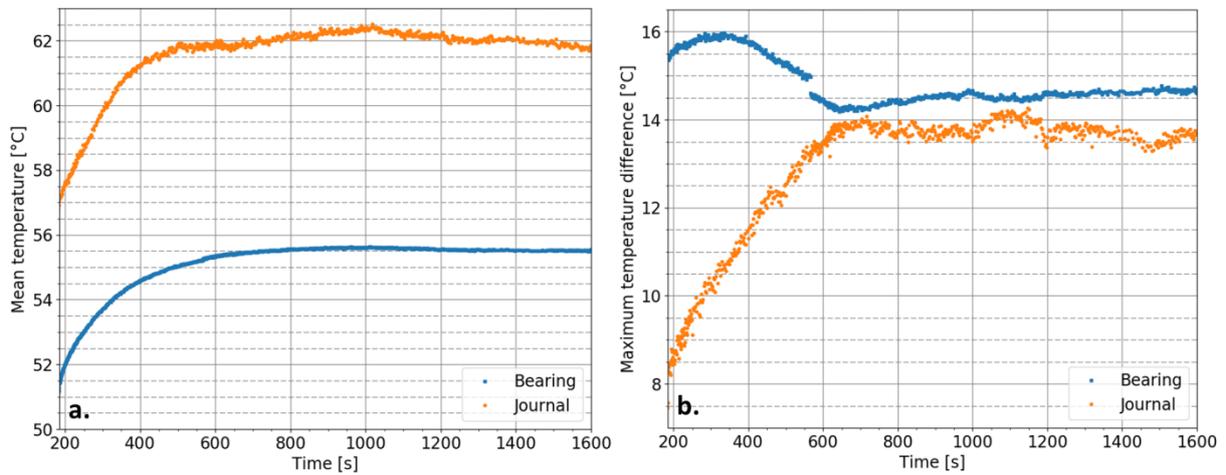

Figure 18 : Journal and bearing average (a) and maximum temperature difference (b) measured for the long rotor at 6.6 krpm, stable response of test 1

**Unstable thermal response of the long rotor (test 2)**

The same test at 6.6 krpm was repeated but the start-up duration was shortened to 80 s. The results are depicted in Figure 19 to Figure 22.

Figure 19 depicts the synchronous amplitudes and phases recorded at 6.6 krpm. The amplitudes increase quite rapidly and the rotor had to be coasted-down after 23 s (at t=103 s from start-up in Figure 19) to avoid contacts. This enlightens an unstable synchronous vibration.

The X direction amplitudes and phases recorded in the NDE plane versus rotor speed are depicted in Figure 20. The amplitudes show a clear hysteresis.

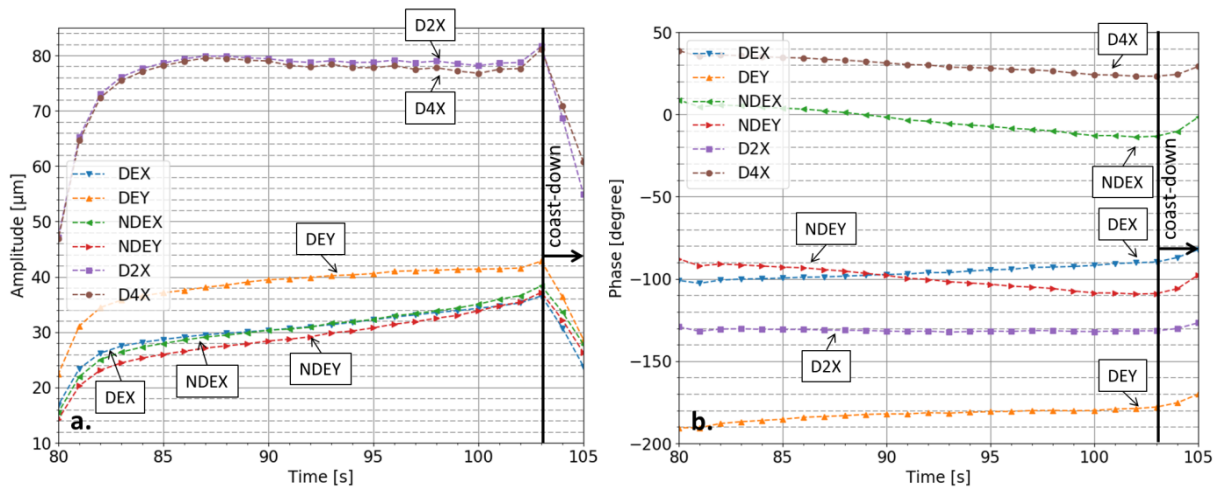

Figure 19 : Synchronous amplitudes (a) and phases (b) measured for the long rotor at 6.6 krpm, unstable response of test 2

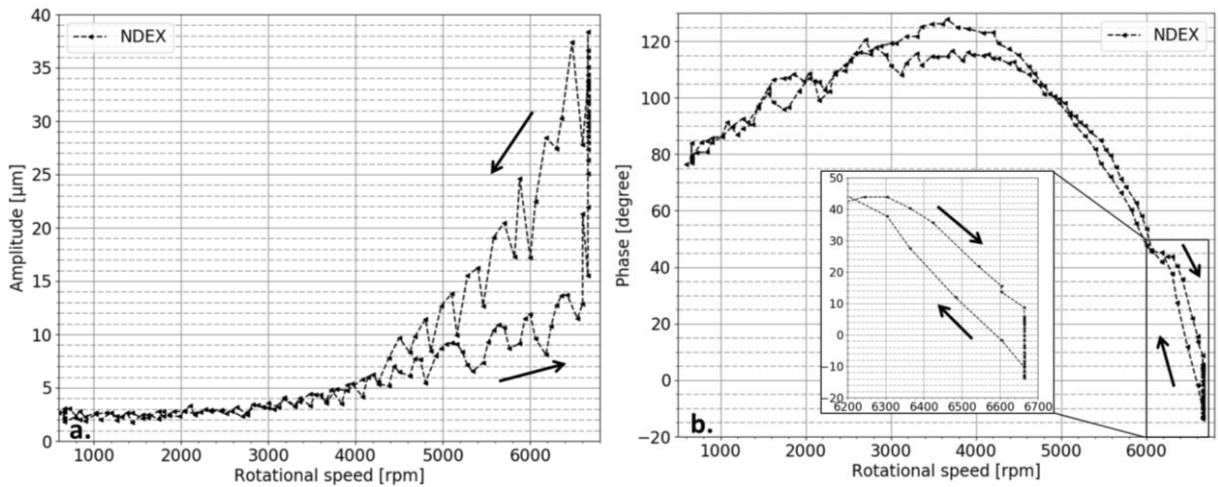

Figure 20 : Synchronous amplitudes (a) and phases (b) at NDEX recorded during start-up and coast-down, unstable response of test 2

The polar plot of the vibration vectors are depicted in Figure 21. Compared to the results depicted in Figure 17, negative phase shifts (i.e. opposite to the rotation speed) were added to the vibration vectors. Moreover, all the vibration vectors depicted in Figure 17 were turning in a direction opposite to the rotation speed, but the results presented in Figure 21 show vibration vectors turning progressively counter-clockwise, in the direction of the rotation speed:

- The X vibration vector of disk 2 shows almost no rotation while the amplitude increases from 50 µm to 85 µm
- The vibration vectors measured at the DE plan rotate in the direction of the rotation speed,
- The vibration vectors measured at the NDE plane and the X vibration vector of disk 4 have a less pronounced rotation opposite to the rotation speed. The vibration vector of disk 4 (overhung mounted) shows also no phase change while increasing from 45 µm to 80 µm, then turns in the direction opposite to the rotation speed while the amplitude increases slightly and then starts turning in the direction of the rotation speed.

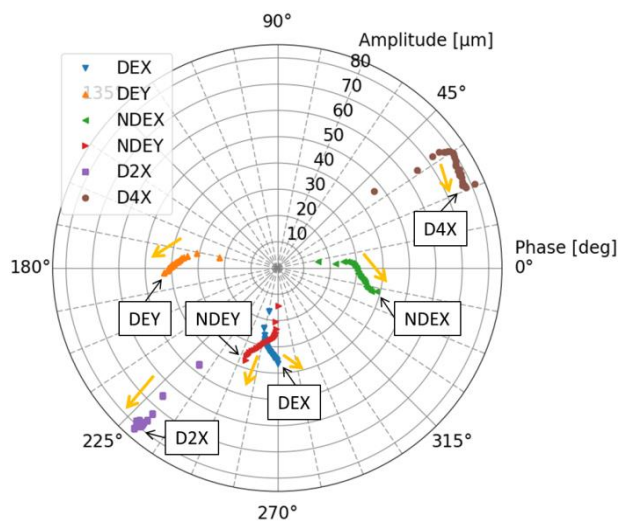

Figure 21 : Polar plot of the vibration vector measured for the long rotor at 6.6 krpm, unstable response of test 2

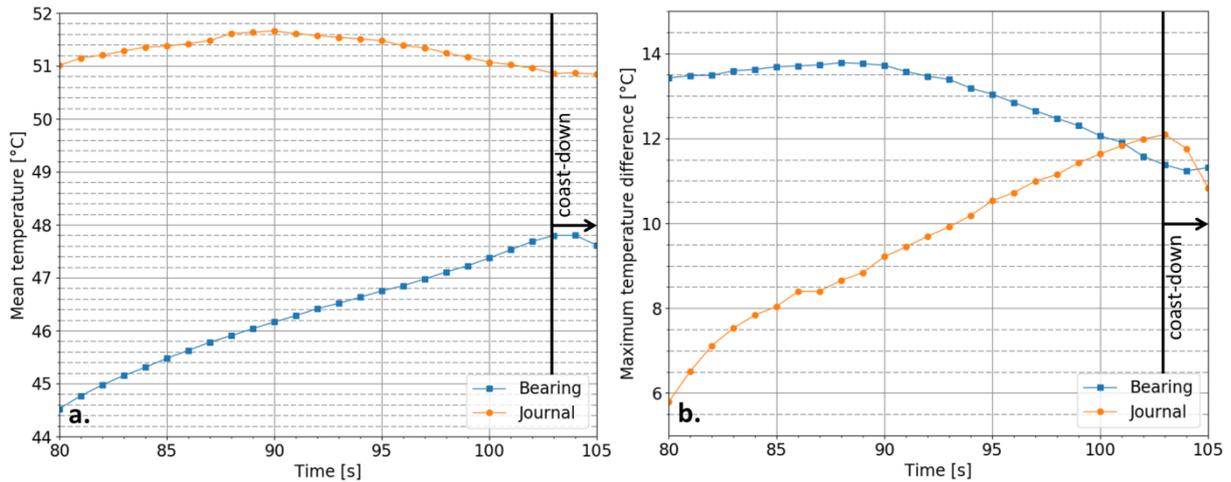

Figure 22 : Journal and bearing average (a) and maximum temperature difference (b) measured for the long rotor at 6.6 krpm, unstable response of test 2

The average and the maximum temperature differences extracted from these measurements are depicted in Figure 22. The journal and the bearing temperatures show different trends while the rotation speed is held constant at 6.6 krpm. The average temperature of the journal is constant or slightly decreases while the bearing average temperature continues to increase. The trends are reversed for the maximum temperature differences. The maximum temperature difference of the bearing is constant and then decreases from 14°C to 12°C. Meanwhile the maximum temperature difference of the journal has a pronounced increase from 6°C to 12°C. These results are due to the increase of the vibration amplitude that makes the bearing temperature more homogenous while the journal temperature is becoming more inhomogeneous. The rapid increase of the journal temperature difference is correlated with the increase of the vibration amplitude.

**Unstable thermal response of the long rotor with contacts (test 3)**

A third test was performed by accelerating again the rotor at 6.6 krpm in 80 s. The synchronous amplitude increased again but the velocity was kept constant more than 23 sec and a light contact occurred between t=103…104 s. The rotor was then coasted-down at t=104 s. The effect of the journal bearing contact is visible on Figure 23 to Figure 25. Figure 23 depicts the amplitude and the phase of the synchronous vibration vector versus time while Figure 24 depicts the polar plot of these vectors.

As depicted in Figure 23a the amplitudes in the DE and NDE planes increase slowly between 80 s and 103 s. Between 103 s and 104 s, when contact occurs, the amplitudes measured in the DE plane decrease slightly while the amplitudes in the NDE plane increase considerably. The amplitudes of the two heavy disks, D2X and D4X show even larger increase. Amplitude variations measured during contact are accompanied by important phase variations depicted in Figure 23b. The fact that the journal bearing contact enhances the rotation of all the vibration vectors (excepting the DE plane) in the direction of rotation speed is visible in Figure 24.

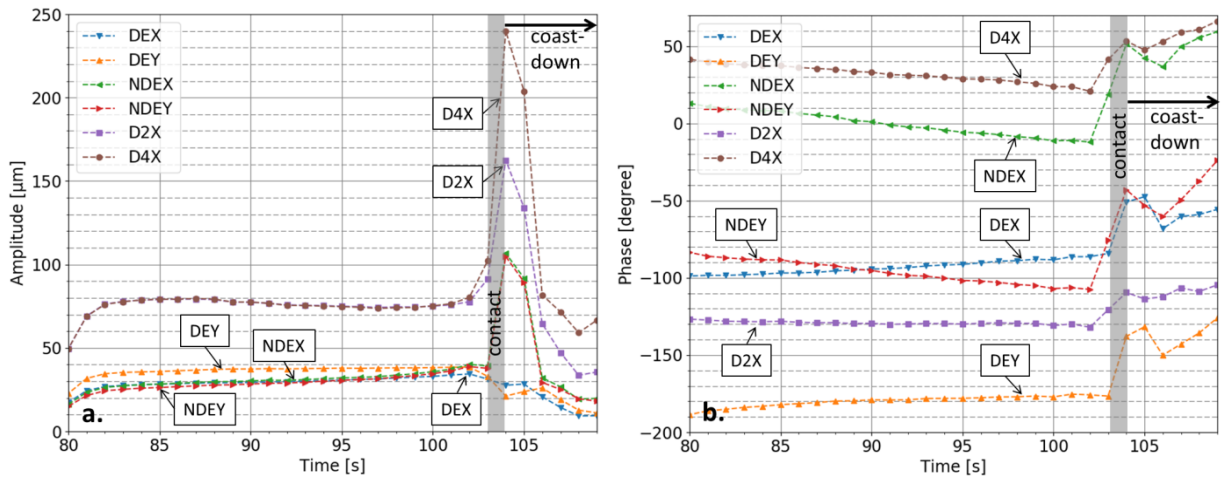

Figure 23 : Synchronous amplitudes (a) and phases (b) measured for the long rotor at 6.6 krpm, test 3 showing contact

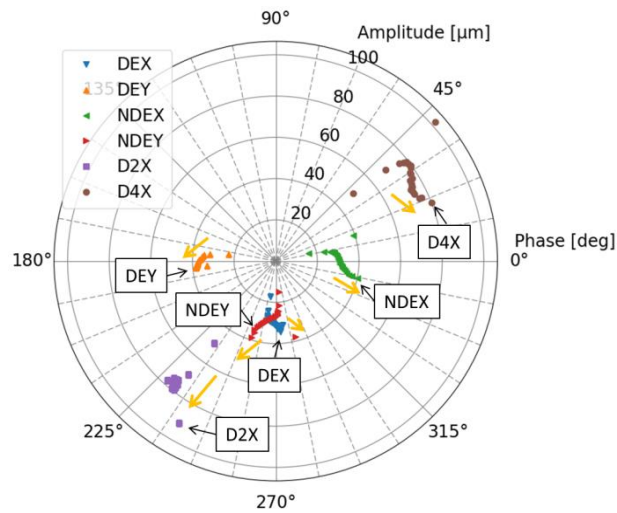

Figure 24 : Polar plot of the vibration vector measured for the long rotor at 6.6 krpm, test 3 showing contact

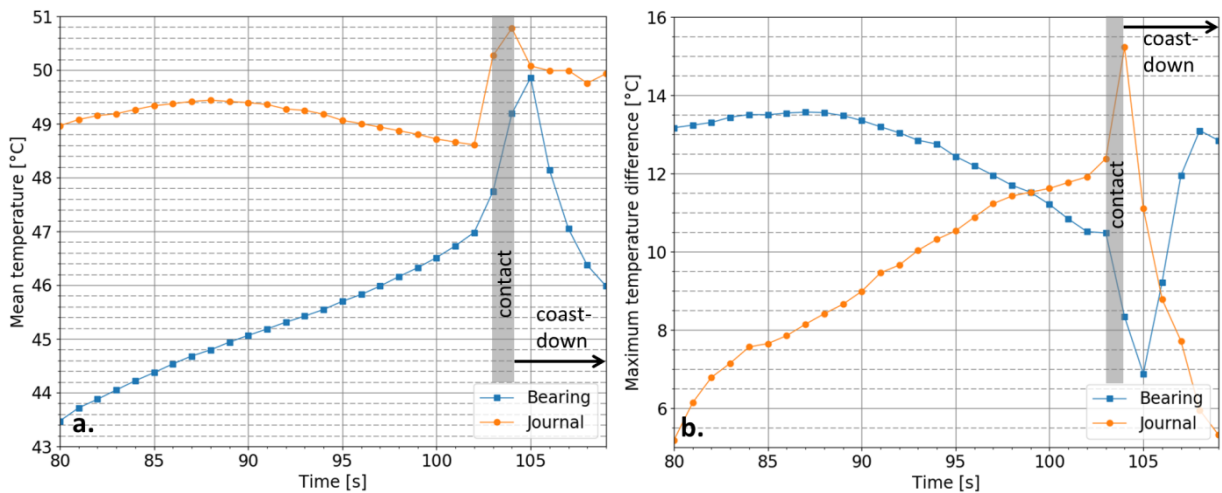

Figure 25 : Journal and bearing average (a) and maximum temperature difference (b) measured for the long rotor at 6.6 krpm, test 3 showing contact

Figure 25 shows the average and the maximum temperature difference. The light contact leads to an increase of the average temperature of the bearing and of the journal but to a decrease of the temperature difference of the bearing. This indicates that the bearing temperatures become more homogeneous following the light contact. The rotor temperature difference shows a notable increase that indicates the development of a strong hot spot.

**Phase lag between the hot spot and high spot of the long rotor**

Figure 26 depicts the phase lag between the hot spot and the high spot following the approach presented for the short rotor. For clarity, error bars were represented only for the phase lags in the journal bearing mid-plane. It should be reminded that error bars were introduced for describing the fact that the phase lag between the hot and the high spot is not constant when the rotor orbit is not centered, nor circular.

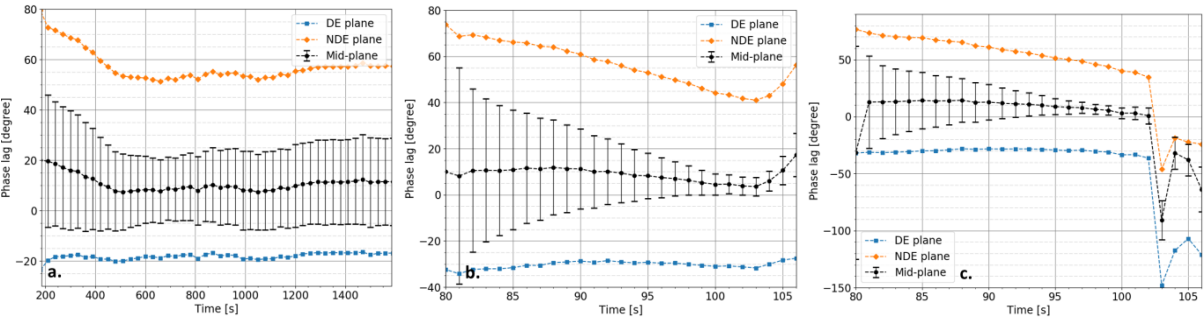

Figure 26 : Phase lag between hot spot and high spot for the long rotor at 6.6 krpm ( a- test 1 stable behavior, b- unstable response, c – test 3 showing contacts)

The phase lags in the DE and NDE planes are different for all three tests: the hot spot precedes the high spot in the DE plane while it lags the high spot in the NDE plane. This result suggests that the rotor is bent in the area of the journal bearing.

For test 1 (stable behavior), the phase lags in the DE and NDE planes become stable after 600 s. The phase lag calculated in the mean plane is comprised between 10° and 15°, significantly lower than for the short rotor.

For test 2 (unstable response), the phase lag in the DE plane has negative, constant values (i.e. the hot spot precedes the high spot) while in the NDE the values are positive but constantly decreasing up to 103 s when the rotor was coasted-down. This means that the main thermal deformation of the rotor occurs in its overhung part. The phase lag calculated in the mean plane is close to zero when the rotor is coasted-down. Zero values of the mid-plane phase lag are also obtained for test 3. This shows that the hot spot and the high spot coincide when contact occurred and it is a natural result. The Morton effect has the tendency to become a Newkirk effect because the heat flux generated by a light journal/bearing contact is larger than any shear heat flux produced by the lubricant.

Also remarkable is the fact that the errors in estimating the phase lag diminish when approaching the journal bearing contact. Indeed, when approaching the contact, the orbits become circular and centered because they describe the bearing circumference. The inherent approximation when calculating the high spot position vanishes.

## Summary and conclusions

The present work introduced the experimental analyses of the thermal unbalance effect induced by a cylindrical journal bearing on a rigid and a flexible rotor. The rotors were tested at similar speeds (7 krpm for the short rotor and 6.6 krpm for the long one) but the operating conditions were very different. The 7 krpm rotation speed of the short rotor was very far from its first flexible critical speed while the 6.6 krpm speed of the long rotor proved to be very close.

All tests showed an important temperature difference on the journal that can bend the rotor. The resulting thermal unbalance had a clear impact on the amplitude of synchronous vibrations. However, the short rotor operated with a stable orbit and showed no instability even for very large vibration amplitudes, close to the radial clearance.

The long rotor showed a similar result when the start-up was 180 s long. When reducing the time length of the start-up to 80 s, the rotor repeatedly showed an unstable response. The synchronous vibration vectors that were initially turning opposite to the rotation speed started to turn progressively in the direction of the rotation speed. This shows that the main part of the thermal unbalance that created the instability is due to the overhung disk. The situation could be different for another rotor configuration.

The phase lag between the hot spot and the high spot decreased progressively for attaining a zero value when contact occurred. The decrease of the phase lag occurred mainly in the NDE plane, close to the overhung disk. This reinforced the conclusion that for the present rotor configuration the thermal bending at the overhung disk is the source of instability.

The results obtained for the long, flexible rotor show that the synchronous thermal instability is triggered not only by the maximum temperature difference of the journal and by the proximity between the rotation speed and the first flexible rotor-bearing critical speed but also by the start-up scenario. Similar results were obtained for seizure effects in cylindrical and tilting pad journal bearings and are reported in [20], [21], [22] and [23]. Different start-up scenarios from zero to operating speed lead to different results and even catastrophic journal bearing seizures.

The present paper is the first part of our investigations of the thermal unbalance effect induced by a journal bearing. The second part [24] is dedicated to theoretical predictions using both full numerical simulations and stability analyses.

# Appendix 1

## Experimental results of the short rotor at 4 krpm

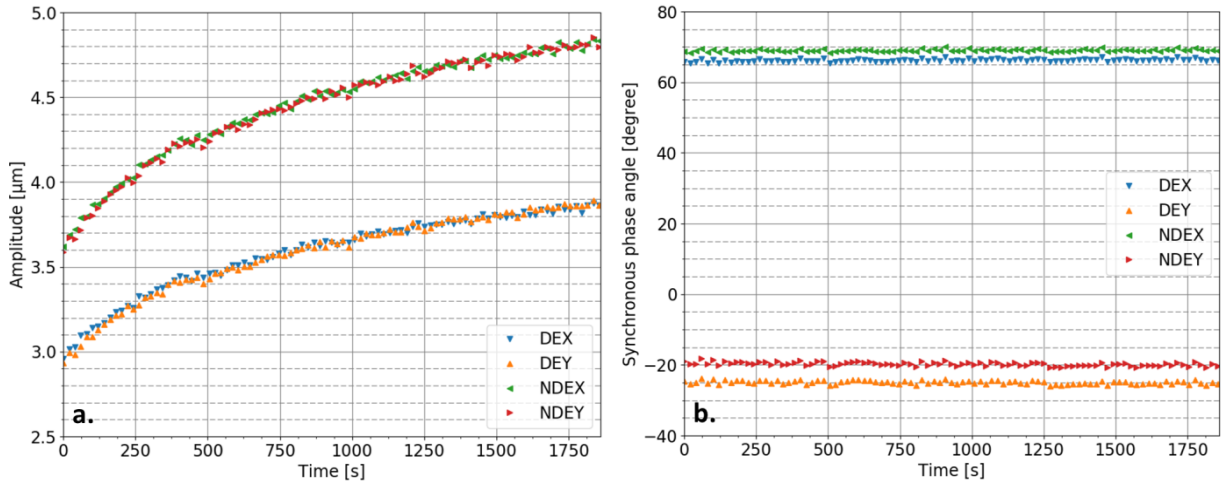

Figure 27 : Synchronous amplitudes (a) and phases (b) measured for the short rotor at 4 krpm

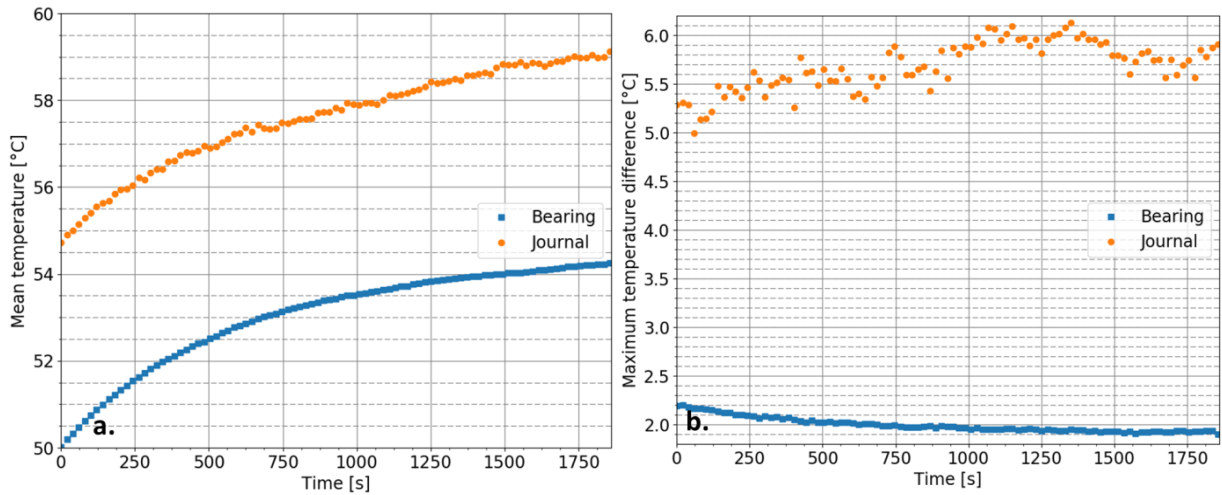

Figure 28 : Average (a) and peak-to-peak temperature differences (b) measured in the mid-plane of the bearing and on the journal for the short rotor at 4 krpm

The first experimental campaign consisted of long tests at 4 krpm. The added unbalance was 60.6gmm@ 180° on the overhung disks. *Figure 27* depicts the measured synchronous amplitudes and phases. The synchronous amplitudes are lower than 5 µm and slowly increasing. However, the phases are clearly constant.

*Figure 28* depicts the average and the maximum (peak-to-peak) temperature differences recorded in the mid-planes of the bearing and of the rotor. The average temperatures are clearly increasing but the peak-to-peak temperature differences tend to constant values. They are not stabilized but it is unlikely that they will trigger a thermal unbalance effect.

# Appendix 2

## Repeatability of the tests with short rotor

Seven tests were conducted with the short rotor (5x2h, 1x4h, 1x6h). The repeatability of the measurements can be estimated from Figure 29 and Figure 30. Figure 29 depicts the circumferentially averaged temperature and the maximum temperature difference on the bearing and on the journal obtained from the seven tests (mean values and the standard deviations). *Figure 30* depicts the mean value and the standard deviation of the synchronous amplitude and phase for the X direction in the NDE plane obtained from the same tests.

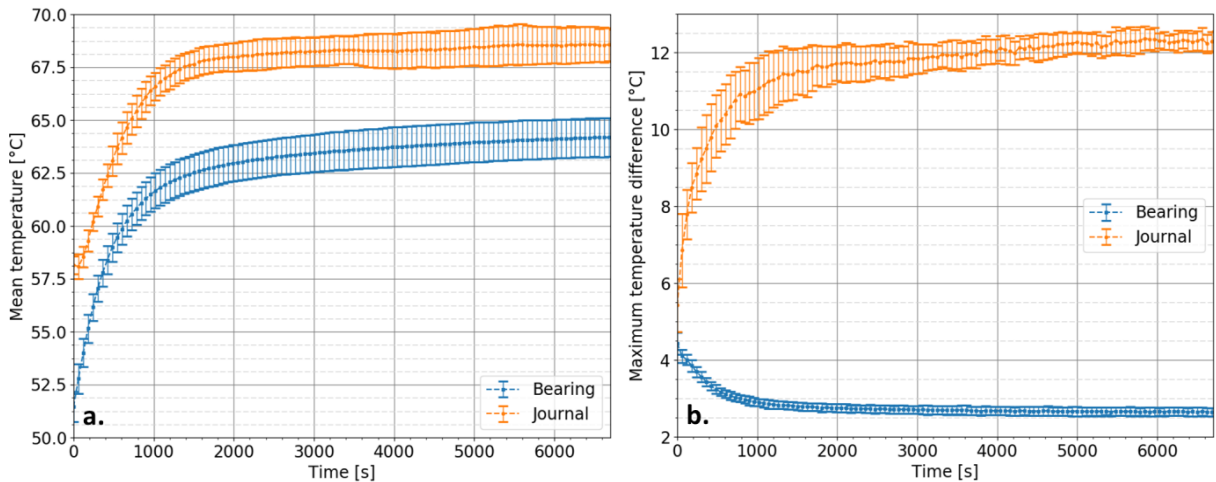

Figure 29 : Journal and bearing average (a) and maximum temperature difference (b) measured for seven tests for the short rotor at 7 krpm

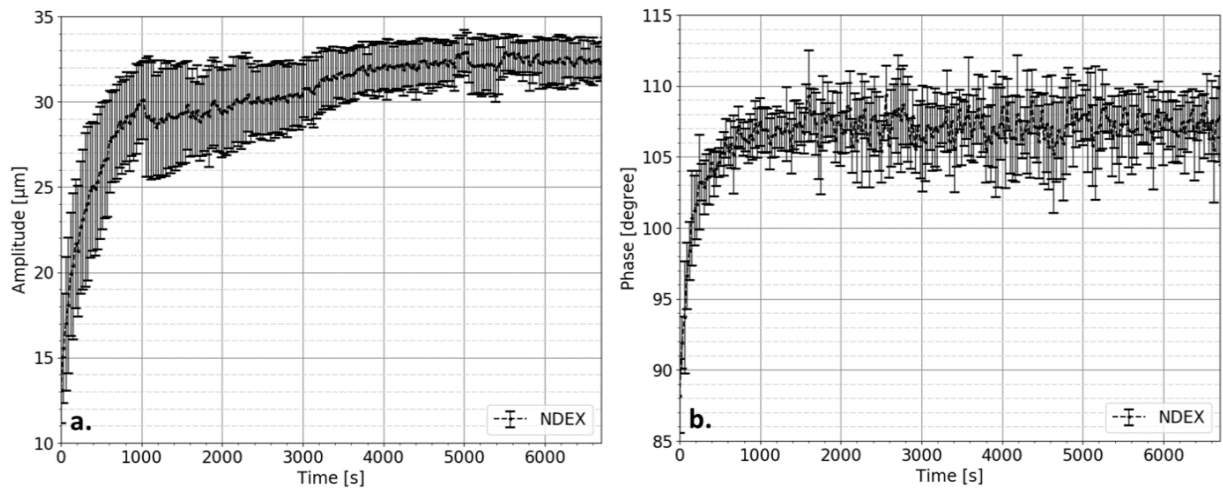

Figure 30 : Synchronous X direction amplitudes (a) and phases (b) in NDE plane for seven tests of the short rotor at 7 krpm

## Acknowledgements

The authors acknowledge the financial support of *Electricité de France*.

Table 1: Disk characteristics for the long rotor

| Disk | Weight [kg] | Length [mm] | Outer diameter [mm] |
|---|---|---|---|
| 1 | 0.2 | 10 | 73.5 |
| 2 | 6.4 | 52 | 149 |
| 3 | 0.4 | 20 | 73.5 |
| 4 | 10.4 | 80 | 152 |